# On mathematical theory of selection: Continuous time population dynamics

Georgiy P. Karev

Lockheed Martin MSD, National Institutes of Health,

Bldg. 38A, Rm. 5N511N, 8600 Rockville Pike, Bethesda, MD 20894, USA

Phone: (301) 451-6722; Fax: (301) 435-7794

E-mail: karev@ncbi.nlm.nih.gov

**Abstract**. Mathematical theory of selection is developed within the frameworks of general models of inhomogeneous populations with continuous time. Methods that allow us to study the distribution dynamics under natural selection and to construct explicit solutions of the models are developed. All statistical characteristics of interest, such as the mean values of the fitness or any trait can be computed effectively, and the results depend in a crucial way on the initial distribution.

The developed theory provides an effective method for solving selection systems; it reduces the initial complex model to a special system of ordinary differential equations (the escort system). Applications of the method to the Price equations are given; the solutions of some particular inhomogeneous Malthusian, Ricker and logistic-like models used but not solved in the literature are derived in explicit form.

# 1. Background

The problem of construction of a formal general theory of selection was clearly formulated in ([30, 32]). The Haldane principle [12], the Fisher Fundamental theorem of natural selection (FTNS) [4], the covariance equation ([23, 34]) and the Price equation [30, 31] were the first outstanding contributions to the future theory. The covariance equation and the FTNS are particular cases of the Price equation (see [27]). Another general approach to the formal theory of selection, models with inheritance, was developed in the 1970s to the 1980s in the works of S. Semenov, V. Okhonin, A. Gorban, et al. (see [8] for references). The approach was based on an abstract version of the so-called replicator equation [13, 36]. The Haldane principle and the Gause competition principle [6] were proven and explored as mathematical assertions within the frameworks of the developed formalism. Recently Grafen [10] also derived the Price equations and the FTNS in a very general form. Both theories are interesting and promising but, perhaps, too abstract for most biological applications.

In this paper we develop an approach to general selection systems that can be applied directly to many mathematical models. We study the evolution of system distributions and obtain the main results in the explicit form. We show that knowing the initial distribution allows us to predict the system dynamics indefinitely and, in particular, to resolve the problem of "dynamical insufficiency" of the Price equations. The developed methods are applicable to a wide class of inhomogeneous population models. We show that the initial complex model can be reduced to the "escort system" of ODEs for auxiliary variables and then solved effectively.

The paper is organized as follows. The master model for selection systems is introduced in s.2. Selection models with self-regulated fitness are explored in s.3, which also contains the main mathematical results, Theorem 1 and its Corollary. S.4.1 contains applications of the theory to the Price equation; the algorithm for solving the selection systems and corresponding replicator equations is described in s.4.2; evolution of particular distributions of biological interest governed by selection over a single or several traits is studied in s.4.3. Explicit solutions of some inhomogeneous Malthusian and logistic-like models used in literature are derived in s.5

# 2. Master model for selection systems

Let us consider a general model of inhomogeneous population, in which every individual is characterized by a vector-parameter $(a_1,...a_n) = \mathbf{a}$ that takes values from set $A$. The parameter $\mathbf{a}$ specifies an individual's inherited invariant properties and may vary from one individual to another, such that the population is non-uniform. Any changes of the parameter distribution with time are caused only by variation of the population structure.

The set of all individuals with a given value of the vector-parameter $\mathbf{a}$ in the population is called $\mathbf{a}$-clone. Let $l(t,\mathbf{a})$ be the density of the population at the moment $t$; informally, $l(t,\mathbf{a})$ is the number of individuals in the $\mathbf{a}$-clone. Let us denote $F(t,\mathbf{a})$ the per capita reproduction rate (Malthusian fitness) of the $\mathbf{a}$-clone at the moment $t$. We suppose that the reproduction rate of every $\mathbf{a}$-clone does not depend on other clones, but can depend on $\mathbf{a}$ and on the "environment" at $t$ moment, which, in turn, may depend on the total population size $N(t)$ and other population characteristics. These quantities evolve with time, however, for a given set of initial conditions have specific values at each point in time for each value of $\mathbf{a}$. The exact form of the reproduction rate will be specified in s. 3.

If we assume the overlapping generations and smoothness of $l(t,\mathbf{a})$ in $t$ for each $\mathbf{a} \in A$, then the population dynamics can be described by the following master model:

$$dl(t,\mathbf{a})/dt = l(t,\mathbf{a})F(t,\mathbf{a}), \qquad (2.1)$$

$$N(t) = \int_A l(t,\mathbf{a})d\mathbf{a},$$

$$P(t,\mathbf{a}) = l(t,\mathbf{a})/N(t)$$

where $P(t,\mathbf{a})$ is the probability density function (pdf) at $t$ instant. The initial pdf $P(0,\mathbf{a})$ and the initial population size $N(0)$ are assumed to be given. Equations (2.1) comprise the formal (after G. Price) selection system with continuous time.

In what follows any characteristic that is inherent to the individual, is fixed at the very beginning of the individual life and does not change with time we refer to as a trait. The selection system describes a closed population of individuals each of which is characterized by a set of qualitative traits; the values of these traits determine the reproduction rate of the individual. It is supposed that the mean values of the traits are the only information that is known about the entire population. The dynamics of the joint distribution of these traits depending on the initial distribution and on correlations between the traits is the main problem of interest.

Hereinafter we use the notation $E^t[f] = \int_A f(\mathbf{a}) P_t(\mathbf{a}) d\mathbf{a}$. It is known (and can be easily proven) that the population size $N(t)$ satisfies the equation

$$dN/dt = NE^t[F] \tag{2.2}$$

and the pdf $P(t,\mathbf{a})$ solves the replicator equation

$$dP(t,\mathbf{a})/dt = P(t,\mathbf{a})(F(t,\mathbf{a}) - E^t[F]). \tag{2.3}$$

The problems (2.1) and (2.2), (2.3) are equivalent.

Taking into account that any smooth function $F(t,\mathbf{a})$ can be well approximated by a finite sum of the form $\sum_i f_i(t)\varphi_i(\mathbf{a})$, we will suppose that the reproduction rate is of the form

$$F(t,\mathbf{a}) = \sum_{i=1}^n f_i(t)\varphi_i(\mathbf{a}). \tag{2.4}$$

Biologically it means that we consider the individual fitness that depends on a given finite set of traits, or "predictors" (see [5, 11, 21], etc.). The function $\varphi_i(\mathbf{a})$ in (2.4) may describe quantitative contribution of a particular *i*-th trait to the total fitness and then $f_i(t)$ are corresponding coefficients of multiple regression. In more sophisticated models, the functions $f_i(t)$ describe relative importance of the trait contributions depending on the environment, population size, etc.

The functions $f_i(t)$ can be known explicitly at any time moment but it is not the case for most realistic models, which accounts for self-limitations of the population growth. For example, even the simple inhomogeneous logistic model with $F(t,\mathbf{a}) = \varphi(\mathbf{a})(1 - N_t/B)$ where constant $B$ is the carrying capacity, does not satisfy this condition because the current population size $N_t$ is unknown a priori. Therefore we should investigate a class of models (2.1), (2.4) where $f_i$ are the functions of the total population size and other population characteristics (see s.3), which are not given and should be computed at every time moment. We show in this paper that this non-trivial problem can be solved effectively.

**3. Self- regulated selection system**

Suppose that the individual reproduction rate can depend on some integral characteristics of the system; we account for extensive characteristics, which depend on the total size of the system (as in most population models) and intensive characteristics, which do not depend on the total size but only on the population frequencies (as in most genetic models). We consider the intensive characteristics in the form

$$H(t) = \int_A h(\mathbf{a})P(t,\mathbf{a})d\mathbf{a} = E^t[h] \qquad (3.1)$$

and the extensive characteristics in the form

$$G(t) = \int_A g(\mathbf{a})l(t,\mathbf{a})d\mathbf{a} = N(t)E^t[g], \qquad (3.2)$$

where $g, h$ are appropriate weight functions. Both expressions (3.1) and (3.2) are known also as "generalized variables" or "regulating functionals"; we will refer to them as "regulators" for brevity. The total system size $N(t)$ is also a regulator (3.2) at $g(\mathbf{a}) \equiv 1$ and is of a special importance.

Suppose that the fitness of every individual is determined by a given set of traits; suppose also that the reproduction rate of every **a**-clone may depend on the size and frequency of other clones only through the regulators. Then we obtain the following general version of the master model:

$$dl(t,\mathbf{a})/dt = l(t,\mathbf{a})F(t,\mathbf{a}), \qquad (3.3)$$

$$F(t,\mathbf{a}) = \sum_{i=1}^{n} u_i(t,G_i)\varphi_i(\mathbf{a}) + \sum_{k=1}^{m} v_k(t,H_k)\psi_k(\mathbf{a}),$$

$$P(t,\mathbf{a}) = l(t,\mathbf{a})/N(t)$$

where $G_i$, $H_k$ are the regulators, $u_i, v_k$ are given functions. The initial pdf $P(0,\mathbf{a})$ and the initial population size $N(0)$ are supposed to be given. In model (3.3) with *self-regulated fitness* the regulators and hence the reproduction rate $F(t,\mathbf{a})$ are not given as explicit functions of time but should be computed employing the current pdf $P(t,\mathbf{a})$ at each time moment.

Let us now formulate the basic theorem for model (3.1)-(3.3). Let $\Phi(r;\boldsymbol{\lambda},\boldsymbol{\delta})$ be the *generating functional*

$$\Phi(r;\boldsymbol{\lambda},\boldsymbol{\delta}) = \int_A r(\mathbf{a})\exp(\sum_{i=1}^{n}\lambda_i\varphi_i(\mathbf{a}) + \sum_{k=1}^{m}\delta_k\psi_k(\mathbf{a}))P(0,\mathbf{a})d\mathbf{a} \qquad (3.4)$$

where $\boldsymbol{\lambda} = (\lambda_1,...\lambda_n)$, $\boldsymbol{\delta} = (\delta_1,...\delta_m)$ and $r(\mathbf{a})$ is a measurable function on $A$.

Define *auxiliary variables* as a solution of the *escort system* of differential equations:

$$dq_i/dt = u_i(t, G_i^*(t)), q_i(0) = 0, i = 1,...n, \quad (3.5)$$

$$ds_k/dt = v_k(t, H_k^*(t)), s_k(0) = 0, k = 1,...m$$

where

$$N^*(t) = N(0)\Phi(1; \mathbf{q}(t), \mathbf{s}(t)), \quad (3.6)$$

$$G_i^*(t) = N(0)\Phi(g_i; \mathbf{q}(t), \mathbf{s}(t)),$$

$$H_i^*(t) = \Phi(h_i; \mathbf{q}(t), \mathbf{s}(t))/\Phi(1; \mathbf{q}(t), \mathbf{s}(t)).$$

Let us denote

$$K_t(\mathbf{a}) = \exp(\sum_{i=1}^{n} q_i(t)\varphi_i(\mathbf{a}) + \sum_{k=1}^{m} s_k(t)\psi_k(\mathbf{a})). \quad (3.7)$$

The function $K_t(\mathbf{a})$ is the reproduction coefficient for the time interval [0,$t$), or *t*-fitness for short (see formula (3.8) below). Notice that $E^0[K_t] = \Phi(1; \mathbf{q}(t), \mathbf{s}(t))$.

The following main theorem reduces model (3.3) to a Cauchy problem for the escort system.

**Theorem 1.** *Let* $0 < T \leq \infty$ *be the maximal value of t such that Cauchy problem* (3.5), (3.6) *has a unique global solution* $\{\mathbf{q}(t), \mathbf{s}(t)\}$ *at* $t \in [0, T)$. *Then the functions*

$$l(t, \mathbf{a}) = l(0, \mathbf{a})K_t(\mathbf{a}), \quad (3.8)$$

$$N(t) = N(0)\Phi(1; \mathbf{q}(t), \mathbf{s}(t)), \quad (3.9)$$

$$G_i(t) = N(0)\Phi(g_i; \mathbf{q}(t), \mathbf{s}(t)), \quad (3.10)$$

$$H_i(t) = \Phi(h_i; \mathbf{q}(t), \mathbf{s}(t))/\Phi(1; \mathbf{q}(t), \mathbf{s}(t)) \quad (3.11)$$

*satisfy system* (3.1)-(3.3) *at* $t \in [0, T)$.

*Conversely, if* $l(t, \mathbf{a})$ *solves system* (3.3) *at* $t \in [0, T)$ *so that* $N(t) = \int_A l(t, \mathbf{a}) d\mathbf{a}$, $H_k(t) = \int_A h_k(\mathbf{a}) l(t, \mathbf{a}) d\mathbf{a}/N(t)$, $G_i(t) = \int_A g_i(\mathbf{a}) l(t, \mathbf{a}) d\mathbf{a}$, *then Cauchy problem* (3.5), (3.6) *has a global solution* $\{\mathbf{q}(t), \mathbf{s}(t)\}$ *at* $t \in [0, T)$ *and the functions* $l(t, \mathbf{a})$, $N(t)$, $G_i(t)$, $H_k(t)$ *can be written in the form* (3.8)-(3.11).

The proof of Theorem 1 is given in Appendix.

**Corollary**. The current system distribution

$$P(t, \mathbf{a}) = P(0, \mathbf{a})K_t(\mathbf{a})/E^0[K_t]. \quad (3.12)$$

The last formula is the central result of the theory, which gives the solution of replicator equation (2.3). Theorem 1 gives an effective algorithm for investigation of the selection systems (see s.6). It allows us to define and compute the total population size and the values of all regulators at any time moment. After that, the reproduction rate $F(t,\mathbf{a})$ in model (3.3) can be considered as a known function of time. In this particular case the theory of selection systems is rather simple (see, e.g., [17]) and all its results can now be applied to self-regulated model (3.3).

**4. Dynamics of the system distribution**

4.1. *The Haldane' principle and the Price equation*

It is known that any stationary distribution of a system with inheritance, and of system (2.1) in particular should be concentrated in the set of points of global maximum of the average reproduction rate on the support of initial distribution; this maximal value must be equal to 0, otherwise the limit distributions cannot exist. This version of the Haldane extreme principle was established in [35, 7, 8]. Similar "selection principle" was proven in [28], s.2.1, 2.2 for logistic-like models. The Haldane principle predicts the behavior of selection systems "at infinity" if the limit distribution exists and is stable. Note that the last condition is not necessarily fulfilled even for the commonly used models; the simplest examples are inhomogeneous Malthusian and some logistic models, see s.5.

On the other end of the time scale, the Price equation describes the instant change of the mean value of individual characters for any selection system. Within the framework of model (2.1) a character, which may depend on time, can be considered a random variable $z(t,\mathbf{a})$ (formally defined on the probabilistic space $\{A,\mathbf{A},P(t,\mathbf{a})\}$ where $\mathbf{A}$ is a $\sigma$-algebra of Borel subsets of $A$). Then the Price equation ([27, 30-33]) states that

$$dE^t[z_t]/dt = Cov^t[F,z_t] + E^t[dz_t/dt]. \qquad (4.1)$$

It is well known ([3, 5, 22], etc.) that the Price equation is dynamically insufficient, i.e., it cannot be used alone as a propagator of the dynamics of the model forward in time. We can write the Price equation in, perhaps, intuitively more clear integral form that shows the connection between the reproduction coefficient and the selection differential $\Delta_t z = E^t[z_t] - E^0[z_0]$, which is an important characteristic of selection.

Let us define the reproduction coefficient for the time interval $[s, s+t)$ as $k_s^{s+t}(\mathbf{a}) = \exp(\int_s^{s+t} F(u,\mathbf{a})du)$. Denote $k_t(\mathbf{a}) = k_0^t(\mathbf{a})$ for brevity; remark that $k_t(\mathbf{a}) = K_t(\mathbf{a})$ for model (3.3). Then

$$E^{t+s}[z_{t+s}] - E^t[z_t] = Cov^t[z_{t+s}, k_t^{t+s}]/E^t[k_t^{t+s}] + E^t[z_{t+s} - z_t]. \qquad (4.2)$$

In particular,

$$E^s[z_s] = Cov^0[z_s, k_s]/E^0[k_s] + E^0[z_s]. \qquad (4.3)$$

Notice that this version of the Price equation is quite similar to the one for discrete-time models, as opposed to the differential version (4.1). Taking into consideration that $K_t^{t+s}(\mathbf{a}) \approx 1 + sF(\mathbf{a})$ we can see that the integral relation (4.2) turns to differential Price' equation (4.1) as $s \to 0$.

It is worth emphasizing that there exists only one reason for "dynamical insufficiency" of the Price equation: this equation is a mathematical identity. Within the frameworks of master model (2.1) the theory developed above helps resolve the problem of dynamical insufficiency of the Price equation and its particular cases, the covariance equation and the equation of the FTNS. Formula (3.12) allows us to compute the mean value of any character in any time moment *if the initial pdf is known*; in this sense, the following proposition gives the "solution" of the Price equation.

**Proposition 1.** *Given the initial distribution, the solution of the Price equation is given by the formulas*

i) $E^t[z_t] = E^0[z_t k_t]/E^0[k_t]$ *for model* (2.1);

ii) $E^t[z_t] = \Phi(z_t; \mathbf{q}(t), \mathbf{s}(t))/\Phi(1; \mathbf{q}(t), \mathbf{s}(t))$ *for self-regulated model* (3.3) *where the auxiliary variables* $\mathbf{q}(t), \mathbf{s}(t)$ *solve the escort ODE system* (3.5).

### 4.2. *How to study selection systems and the dynamics of their distribution?*

Not only asymptotic behavior, but also the current dynamics of the population distribution during protracted but finite time intervals are of interest and, perhaps, of primary importance for applications.

The developed theory yields an effective algorithm to analyze selection systems, which reduces complex self-regulated selection system to the escort system of non-autonomic ODEs. Let us summarize the main steps of the algorithm.

We study a model of selection system in the form (3.3). In applications, the functions $\varphi_i(\mathbf{a}), \psi_k(\mathbf{a})$ can be interpreted as traits that characterize an individual; $u_i(t, G_i)$ and $v_k(t, H_k)$ describe the contribution of the corresponding traits to the individual fitness at moment $t$. The following steps should be performed for solving a particular selection system:

1) composing generating functional (3.4);

2) solving escort system of ODE (3.5).

3) After that, the solution to the selection system $l(t, \mathbf{a})$, the population size $N(t)$, and the values of regulators at $t$ moment are given by formulas (3.7)-(3.11);

4) the current system distribution is given by formula (3.12), which allows us to compute all statistical characteristics of interest for a self-regulated selection system.

Generating functional (3.4) is the main tool of the suggested approach; it may be difficult to compute it in general case. Remark, that we in fact do not need to know this functional for arbitrary function $r$ but only for the functions $g_k$, $k = 1,...m$, and $h_i, i = 1,...n$, see (4.1) and (4.2). So, instead of generating functional (3.4) we can use the *moment generating function* of the initial joint distribution of the set of random variables $\{\varphi_i, g_i, \psi_k, h_k\}$,

$$M_0(\boldsymbol{\lambda}, \boldsymbol{\delta}, \boldsymbol{\gamma}, \boldsymbol{\eta}) = \int_A \exp(\sum_{i=1}^n (\lambda_i \varphi_i(\mathbf{a}) + \gamma_i g_i(\mathbf{a})) + \sum_{k=1}^m (\delta_k \psi_k(\mathbf{a}) + \eta_k h_k(\mathbf{a})) P(0, \mathbf{a}) d\mathbf{a}. \quad (4.4)$$

Indeed,

$$\Phi(g_i; \mathbf{q}(t), \mathbf{s}(t)) = \int_A g_i(\mathbf{a}) \exp(\sum_{i=1}^n q_i(t) \varphi_i(\mathbf{a}) + \sum_{k=1}^m s_k(t) \psi_k(\mathbf{a})) P(0, \mathbf{a}) d\mathbf{a} =$$

$$\frac{\partial}{\partial \gamma_i} M_0(\mathbf{q}(t), \mathbf{s}(t), \boldsymbol{\gamma}, \boldsymbol{\eta})\bigg|_{\gamma=\eta=0},$$

$$\Phi(h_k; \mathbf{q}(t), \mathbf{s}(t)) = \frac{\partial}{\partial \nu_k} M_0(\mathbf{q}(t), \mathbf{s}(t), \boldsymbol{\gamma}, \boldsymbol{\eta})\bigg|_{\gamma=\eta=0}.$$

The general method is simplified in an important case of the reproduction rate $F(t, \mathbf{a}) = \sum_{i=1}^n f_i(t, S_i) \phi_i(\mathbf{a})$ with regulators $S$ of the forms $N(t), E^t[\phi_i], N(t)E^t[\phi_i]$ only. In this case we can use the moment generating function of the joint initial distribution of the variables $\{\phi_i\}$ only, $M_0(\boldsymbol{\lambda}) = E^0[\exp(\sum_{i=1}^n \lambda_i \phi_i)]$, instead of general mgf (4.4). The escort system reads

$$dq_i / dt = f_i(t, S_i(t)), q_i(0) = 0, i = 1,...n \qquad (4.5)$$

where $S_i(t)$ (having the form $N(t)$, $E^t[\phi_i]$, $N(t)E^t[\phi_i]$) are defined by the formulas

$$N(t) = N(0)M_0(\mathbf{q}(t)), \qquad (4.6)$$

$$E^t[\phi_k] = \partial_k \ln M_0(\mathbf{q}(t)).$$

Here we denoted $\partial_k M_0(\boldsymbol{\lambda}) = \partial M_0(\boldsymbol{\lambda})/\partial \lambda_k$ for brevity. This simplified version of the algorithm works for many models, see s.5. It is important that the moment generating functions are known for most discrete and continuous distributions used in biological applications.

4.3. *Dynamics of particular distributions*

Let us formulate the following assertions that capture the dynamics of the system distribution for some probability distributions of biological interest. A selection system, whose evolution is governed by selection over a single trait, $\varphi(\mathbf{a})$, is the simplest but widely spread and important case. The reproduction rate may depend on regulators (3.1), (3.2); for simplicity, suppose that it depends only on extensive regulators like the total population size. Then the system is of the form

$$dl(t, \mathbf{a})/dt = l(t, \mathbf{a})F(t, \mathbf{a}), \qquad (4.7)$$

$$F(t, \mathbf{a}) = u_0(t, G_0) + u_1(t, G_1)\varphi(\mathbf{a}),$$

$$G_i(t) = \int_A g_i(\mathbf{a})l(t, \mathbf{a})d\mathbf{a}.$$

The regulating functional for this system

$$\Phi(r; \lambda_0, \lambda_1) = \exp(\lambda_0)\int_A r(\mathbf{a})\exp(\lambda_1 \varphi(\mathbf{a}))P(0, \mathbf{a})d\mathbf{a}. \qquad (4.8)$$

Define the auxiliary variables $q_0(t), q_1(t)$ by the escort system:

$$dq_0/dt = u_0(t, N(0)\Phi(g_0; q_0, q_1)),$$
$$dq_1/dt = u_1(t, N(0)\Phi(g_1; q_0, q_1)), \qquad (4.9)$$
$$q_i(0) = 0, i = 1,2.$$

**Proposition 2.**

*Consider model (4.7) and assume that the initial distribution of the trait $\varphi(\mathbf{a})$ is*

(i) *normal with mean* $m_0$ *and variance* $\sigma_0^2$. *Then the trait distribution will also be normal at any t with mean* $m_t = m_0 + \sigma_0^2 q_1(t)$ *and with the same variance* $\sigma_0^2$;

(ii) *Poisson with mean* $m_0$. *Then the trait distribution will also be Poisson at any t with the mean* $m_t = m_0 \exp(q_1(t))$;

(iii) $\Gamma$ *-distribution with the coefficients* $k, a, \eta$, *i.e.* $P_0(\varphi = x) = a^k (x-\eta)^{k-1} \exp(-(x-\eta)a)/\Gamma(k)$, *where* $k, a > 0, -\infty < \eta < \infty$, $x \geq \eta$; $\Gamma(k)$ *is the* $\Gamma$ *- function.*

*Define* $T^* = \inf(t : q_1(t) = a)$, *if such t exists, otherwise* $T^* = T$. *Then the trait* $\varphi$ *will be* $\Gamma$ *- distributed at any time moment* $t < T^*$ *with coefficients* $a - q_1(t), k, \eta k$, *such that* $E^t[\varphi] = \eta + k/(a - q_1(t))$, $\sigma_t^2 = k/(a - q_1(t))^2$.

The list of practically implemented distributions can be extended.

Now let us consider model (4.7) with many traits when

$$F(t, \mathbf{a}) = \sum_{i=1}^{n} u_i(t, G_i) \varphi_i(\mathbf{a}). \tag{4.10}$$

The regulating functional for system (4.7), (4.10)

$$\Phi(r; \lambda) = \int_A r(\mathbf{a}) \exp(\sum_{i=1}^{n} \lambda_i \varphi_i(\mathbf{a})) P(0, \mathbf{a}) d\mathbf{a}$$

and the auxiliary variables $q_i(t)$ solve the escort system:

$$dq_i / dt = u_i(t, N(0) \Phi(g_i; \mathbf{q})), \tag{4.11}$$

$q_i(0) = 0, i = 1,...n.$

If initially the traits $\varphi_i(\mathbf{a})$ are independent (as random variables on probabilistic space $\{A, \mathbf{A}, P(0, \mathbf{a})\}$) then they remain independent indefinitely (as random variables on probabilistic space $\{A, \mathbf{A}, P(t, \mathbf{a})\}$ for any $t$) and given the initial mgf their joint mgf can be easily computed at any time moment (see [17]). In reality, the evolution of a system is governed by simultaneous selection over many traits, whose contributions to fitness depend on each other. The evolution of the pdf of the vector $\boldsymbol{\varphi} = (\varphi_1, ... \varphi_n)$ in the general case of correlated traits $\{\varphi_i, i = 1,...n\}$ is of great practical interest. Let us recall some definitions (see [20]).

A random vector $\mathbf{X} = (X_1, \ldots X_n)$ has a multivariate normal distribution with the mean $E\mathbf{X} = \mathbf{m} = (m_1, \ldots m_n)$ and covariance matrix $\mathbf{C} = \{c_{ij}\}, c_{ij} = \mathrm{cov}(X_i, X_j)$ if its mgf is $M(\boldsymbol{\lambda}) = E[\exp(\boldsymbol{\lambda}^T \mathbf{X})] = \exp(\boldsymbol{\lambda}^T \mathbf{m} + 1/2 \boldsymbol{\lambda}^T \mathbf{C} \boldsymbol{\lambda})$.

A random vector $\mathbf{X} = (X_1, \ldots X_n)$ has a multivariate polynomial distribution with parameters $(k; p_1, \ldots p_n)$, if $P(X_1 = m_1, \ldots X_n = m_n) = \frac{k!}{m_1! \ldots m_n!} p_1^{m_1} \ldots p_n^{m_n}$ for $\sum_{i=1}^{n} m_i = k$. The mgf of the polynomial distribution is $M(\boldsymbol{\lambda}) = (\sum_{i=1}^{n} p_i \exp(\lambda_i))^k$.

A general class of *multivariate natural exponential distributions* is especially important for selection systems and their applications. It includes multivariate polynomial, normal, and other distributions as special cases. A random $n$-dimension vector $\mathbf{X} = (X_1, \ldots X_n)$ has multivariate natural exponential distribution (NED) with parameters $\boldsymbol{\theta} = (\theta_1, \ldots \theta_n)$ with respect to the positive measure $\nu$ on $R^n$ if its joint density function is of the form $f(\mathbf{X}) = h(\mathbf{X}) \exp(\mathbf{X}^T \boldsymbol{\theta} - s(\boldsymbol{\theta}))$ where $s(\boldsymbol{\theta})$ is the normalization function. The mgf of NED is

$$M(\boldsymbol{\lambda}) = E_\nu[\exp(\boldsymbol{\lambda}^T \mathbf{X})] = \exp(s(\boldsymbol{\theta} + \boldsymbol{\lambda}) - s(\boldsymbol{\theta})).$$

**Proposition 3**. *Let us assume that at the initial time moment the vector* $\boldsymbol{\varphi} = (\varphi_1, \ldots \varphi_n)$ *has*

i) *multivariate normal distribution with the mean vector* $\mathbf{m}(0)$ *and covariance matrix* $\mathbf{C} = (c_{ij})$. *Then the vector* $\boldsymbol{\varphi}$ *also has the multivariate normal distribution at any moment* $t < T$ *with the same covariance matrix* $\mathbf{C}$ *and the mean vector* $\mathbf{m}(t), m_i(t) = m_i(0) + 1/2 \sum_{k=1}^{n} (c_{ik} + c_{ki}) q_k(t)$ ;

ii) *multivariate polynomial distribution. Then the vector* $\boldsymbol{\varphi}$ *also has the multivariate polynomial distribution at any moment* $t < T$ *with parameters* $(k; p_1(t), \ldots p_n(t))$ *where*

$$p_i(t) = p_i \exp(q_i(t)) / \sum_{j=1}^{n} p_j \exp(q_j(t));$$

iii) *multivariate natural exponential distribution on* $R^n$ *with respect to the Lesbegue measure, with the density function* $f_0(\mathbf{X}) = h(\mathbf{X}) \exp(\mathbf{X}^T \boldsymbol{\theta} - s(\boldsymbol{\theta}))$ *and the mgf* $M(\boldsymbol{\lambda}) = \exp(s(\boldsymbol{\theta} + \boldsymbol{\lambda}) - s(\boldsymbol{\theta}))$. *Then the vector* $\boldsymbol{\varphi}$ *also has the multivariate NED at any* moment $t < T$ with the parameters $\boldsymbol{\theta} + \mathbf{q}(t)$, the density function $f_t(\mathbf{X}) = h(\mathbf{X})\exp(\mathbf{X}^T(\boldsymbol{\theta} + \mathbf{q}(t)) - s(\boldsymbol{\theta} + \mathbf{q}(t)))$ and the moment generating function $M_t(\boldsymbol{\lambda}) = \exp(s(\boldsymbol{\theta} + \boldsymbol{\lambda} + \mathbf{q}(t)) - s(\boldsymbol{\theta} + \mathbf{q}(t)))$.

Proofs of Propositions 2, 3 are similar to that given in [17] for a less general model. Both assertions up to the definition of the auxiliary variables are valid for selection systems whose fitness may depend on some intensive regulators.

## 5. Applications

The simplest selection system is the inhomogeneous Malthusian-like model $dl(t, \mathbf{a})/dt = l(t, \mathbf{a})\varphi(\mathbf{a})$. The function $\varphi(\mathbf{a})$ itself can be often considered as a distributed parameter, and then the model reads $dl(t, x)/dt = xl(t, x)$.

Let $M(\lambda) = \int_A \exp(\lambda x) P_0(x) dx$ be the mgf of the initial distribution of the parameter $x$. Then the solution of the model $l(t, x) = \exp(xt) l(0, x)$ and $N(t) = N_0 M(t)$. The model distribution solves the replicator equation $dP_t(x)/dt = P_t(x)(x - E^t[x])$; its solution is $P_t(x) = P_0(x)\exp(xt)/M(t)$.

Inhomogeneous Malthusian models and their applications to some problems of forest ecology and global demography were studied in [14-16]. It was shown that even this simplest inhomogeneous model possess a variety of solutions depending on the initial distribution, which may have many interesting and even counterintuitive peculiarities. Let us demonstrate some of them on the example of inhomogeneous Malthusian-like model with limiting factors.

**Example 1.**

Principle of limiting factors, according to Liebig [24], states that at any given moment the rate of a process is determined by the factor whose sufficiently small modification produces a change of the rate; it is assumed that similar changes in other factors do not affect the rate (see [29] for mathematical formulation). The principle of limiting factors was actually used in a model of early biological evolution suggested in [39]. Each organism was characterized by the vector **a** where the component $a_i$ is the thermodynamic probability that protein $i$ is in its native conformation. In order to study the connection between molecular evolution and population, the authors suppose that the organism death rate $d$ depends on the stability of its proteins as $d = d_0(1 - \min a_i)$, $d_0 = b/(1-a_0)$, $b$ is the birth rate, $a_0$ is the native state probability of a protein. Hence, neglecting possible mutations (accounted for by the authors in their simulations), the model can be formalized as the system

$$dl(t,\mathbf{a})/dt = l(t,\mathbf{a})d_0(m(\mathbf{a}) - a_0)) \qquad (5.1)$$

where $m(\mathbf{a}) = \min[a_1,...a_n]$. In what follows we let $d_0 = 1$ for simplicity. It was supposed in [39] that the values $a_i$ are independent from each other and have the Boltzmann distribution. We can consider $a_i$ as the $i$-th realization of a random variable with a common pdf $f(a)$. Let $G(a) = \int_0^a f(x)dx$ be the cumulative distribution function. Then, it is well known that the pdf of $\min[a_1,...a_n]$ is equal to $g(a) = n(1-G(a))^{n-1}f(a)$. Equation (5.1) is a version of the inhomogeneous Malthusian equation, which can now be solved explicitly at any given pdf $f(a)$. In particular, if

$$f(a) = \exp(-a/T)/Z, \quad Z = \sum_a \exp(-a/T) \qquad (5.2)$$

is the Boltzmann distribution with $a > 0$, then

$$g(a) = n(1-G(a))^{n-1}f(a) = (n(\sum_{x>a}\exp(-x/T))^{n-1})\exp(-a/T)/\sum_x \exp(-x/T)).$$

For distribution (5.2) with continuous range of values of $a$, $a \in (0,\infty)$, $Z = T$, $1 - G(x) = \exp(-x/T)$ and

$$g(a) = n(\exp(-a(n-1)/T)\exp(-a/T)/T = n/T \exp(-an/T). \qquad (5.3)$$

If $a \in (0,E)$, then $Z = T(1-\exp(-E/T))$, $1 - G(a) = \dfrac{\exp((E-a)/T)-1}{\exp(E/T)-1}$, and

$$g(a) = \frac{n\exp(-a/T)}{T(1-\exp(-E/T))}\left[\frac{1-\exp((E-a)/T)}{1-\exp(E/T)}\right]^{n-1}. \qquad (5.4)$$

Let $M_0(\lambda) = E^0[\exp(\lambda m)]$. For initial distribution (5.3), $M_0(\lambda) = \dfrac{1}{1-\lambda T/n}$. Hence,

$$l(t,\mathbf{a}) = l(0,\mathbf{a})\exp((m(\mathbf{a}) - a_0)t),$$

$$N(t) = N(0)\exp(-a_0 t)\frac{1}{1-tT/n},$$

$$P(t,\mathbf{a}) = P(0,\mathbf{a})\exp(m(\mathbf{a})t)(1-tT/n).$$

At the moment $t_{max} = n/T$ the population „blows up": $N(t)$ and $l(t,\mathbf{a})$ tend to infinity as $t \to t_{max}$. Let us denote $p(t,a) = P(t,\{\mathbf{a}: m(\mathbf{a}) = a\})$. Then at $t < t_{max}$

$$p(t,a) = n/T \exp(-an/T + at)(1 - tT/n) = (n/T - t)\exp(a(t - n/T)).$$

The probability $P(t,\{\mathbf{a}: m(\mathbf{a}) < a\})$ tends to 0 for any finite $a$ as $t \to t_{max}$. Loosely speaking, the total "probability mass" goes to infinity after a finite time interval. So, we should conclude that model (5.1), (5.2) which allow arbitrary large values of the parameter $a$ with nonzero probability have no "physical" sense.

This problem can be eliminated by taking the initial distribution (5.4), which allows only bounded values of the parameter $a$. For pdf (5.4), the integral $M_0(\lambda) = \int_0^E \exp(\lambda x) g(x) dx$ is finite for any $\lambda$; although it cannot be expressed in quadratures we can obtain much information about the system distribution and its dynamics. The current pdf

$$p(t,a) = \frac{n}{T(\exp(E/T) - 1)^n} \exp((E-a)/T)(\exp((E-a)/T) - 1)^{n-1} \frac{\exp(at)}{M_0(t)}$$

where $M_0(t)$ is finite for all $t$. So, the pdf is well defined at any time moment, in contrast to the previous case. The total distribution concentrates with time at the point $a = E$, which provides the maximal reproduction rate. Let us emphasize that the pdf $p(t,a)$ does not depend on the native state probability $a_0$.

Example 1 demonstrates a possibility of the "blowing up" phenomenon, when the total population size goes to infinity after a finite time interval. Similar phenomenon was discovered earlier in models of global demography [16]. One may suppose that the choice of the Malthusian model for population dynamics is to blame for this phenomenon and the problem should disappear if the Malthusian model is replaced by the logistic model. Surprisingly, it is not the case for *inhomogeneous* logistic models. Let us apply our approach to a wide class of generalized inhomogeneous logistic- type models:

$$dl(t,\mathbf{a})/dt = l(t,\mathbf{a})F(t,\mathbf{a}), \tag{5.5}$$

$$F(t,\mathbf{a}) = f_1(t)b(\mathbf{a}) - f_2(t)d(\mathbf{a})$$

where $f_1(t)b(\mathbf{a})$ is the birth rate, $f_2(t)d(\mathbf{a})$ is the death rate, and $f_i$ are functions of the regulators $H(t) = E^t[h]$ or $G(t) = N(t)E^t[g]$. Different versions of equation (5.5) were discussed in numerous works. Theorems of existence and uniqueness and asymptotic behavior of some equations of the form (5.5) were studied in [1, 2, 7, 8, 28]. The method described in s. 4.2 allows us to obtain the solution of equation (5.5) at any instant within the time interval where the global solution of the escort system exists. We will not rewrite the general formulas in terms of equation (5.5); instead, we give the solutions of some particular inhomogeneous logistic equations used (but not solved explicitly) in the literature.

**Example 2.**

The following example of the system with inheritance was considered in [8]:

$dl(t,\mathbf{a})/dt = l(t,\mathbf{a})F(t,\mathbf{a})$,

$F(t,\mathbf{a}) = b(\mathbf{a}) - \int_A m(\mathbf{a})l(t,\mathbf{a})d\mathbf{a} = b(\mathbf{a}) - N(t)E^t[m]$.

It can be interpreted as follows. Let $b(\mathbf{a})$ be the specific birth rate of inherited varieties $\mathbf{a}$. The death rate is determined by the common factor $\int_A m(\mathbf{a})l(t,\mathbf{a})d\mathbf{a}$ where $m(\mathbf{a})$ is the individual contribution of variety $\mathbf{a}$ to this death rate. Using the method summarized above, we are now able to give an explicit solution of this model. Compose the generating functional $\Phi(r;\lambda_1,\lambda_2) = \exp(-\lambda_2)E^0[r\exp(\lambda_1 b)]$. Then the escort system reads

$dq_1/dt = 1, q_1(0) = 0,$ hence $q_1(t) = t$;

$dq_2/dt = N(0)\Phi(m;q_1,q_2) = N(0)\exp(-q_2)E^0[m\exp(tb)], q_2(0) = 0$.

This equation can be easily integrated since $E^0[m\exp(tb)] = f(t)$ is a known function of $t$ at given $P(0,\mathbf{a})$: $\exp(q_2(t)) = 1 + N(0)\int_0^t f(s)ds = 1 + N(0)E^0[\frac{m}{b}(\exp(tb)-1)]$.

The solution to the model

$l(t,\mathbf{a}) = l(0,\mathbf{a})\exp(tb(\mathbf{a}) - q_2(t)) =$

$l(0,\mathbf{a})\exp(tb(\mathbf{a}))/\{1 + N(0)E^0[\frac{m}{b}(\exp(tb)-1)]\}$. (5.6)

Also, $N(t) = N(0)E^0[\exp(tb)]/\{1 + N(0)E^0[\frac{m}{b}(\exp(tb)-1)]\}$, and

$P(t,\mathbf{a}) = l(t,\mathbf{a})/N(t) = P(0,\mathbf{a})\exp(tb(\mathbf{a}))/E^0[\exp(tb)]$.

Remark, that the current pdf $P(t,\mathbf{a})$ does not depend on the death rate. We can observe now an interesting phenomenon, which is impossible for "homogeneous" logistic models. Let us suppose that the birth rate $b(\mathbf{a})$ is $\Gamma$-distributed at the initial moment, i.e. $P_0(b = x) = T^k x^{k-1} \exp(-xT)/\Gamma(k)$, where $k, T > 0$ are constants, $x \geq 0$.

Then $E^0[\exp(tb)] = (1 - t/T)^{-k}$ for $t < T$. So, $P(t, \{b < x\}) < \exp(tx)(1 - t/T)^k \to 0$ as $t \to T$ for any $x$; by words, the frequency of individuals with a finite birth rate vanishes as $t \to T$.

**Example 3.**

There exist a large number of papers devoted to the problem of evolution of altruism. We do not discuss here an interesting and important problem how an altruistic trait can be selected. Instead, using the developed tools, we give an exact solution of a model used in [38]. Consider a population in which each individual possesses a trait $x$ that increases the fitness of everyone in the population (including itself) by an amount $mx$ at a personal cost $-cx$. Then the fitness of the individual is

$$F(t, x) = -cx + mN(t)E^t[x], x > 0, \tag{5.7}$$

which coincides with the fitness of Example 2 up to notation but has the opposite sign. It is interesting that model (5.7) exhibits essentially different dynamical behavior. Using formula (5.6) we can write the model solution as

$$l(t, x) = l(0, x)\exp(-tcx)/\{1 + N(0)m/c(E^0[\exp(-tcx)] - 1\}. \tag{5.8}$$

$$N(t) = N(0)E^0[\exp(-tcx)]/\{1 + N(0)m/cE^0[\exp(-tcx)] - 1\}, \text{ and}$$

$$P(t, x) = P(0, x)\exp(-tcx)/E^0[\exp(-tcx)].$$

As $c > 0$, the model solution exists and is finite indefinitely for any initial distribution of the trait.

**Example** 4.

A microbial population in an environment of an antimicrobial agent was studies in [25]; the suggested model has (up to notation) the form of inhomogeneous Malthusian equation

$$dN/dt = (K - m(C))N \tag{5.9}$$

where $K$ is the physiological growth rate, $m(C)$ is the kill rate induced by the antimicrobial agent, which has concentration $C$. A more accurate version with logistic growth rate was also discussed:

$$dN/dt = KN(1 - N/B) - m(C)N. \tag{5.10}$$

The value of $m(C)$ in comparison to $K$ represents the resistance of microbes to a specific antimicrobial agent with concentration $C$. Population resistance is distributed over a multitude of values. For the Malthusian version of the model, (5.9), the authors derive the equation for the size of an entire population over time and then approximate it using the variance and higher-order cumulants of distribution of the kill rate over a heterogeneous population.

The theory of inhomogeneous Malthusian and logistic equations developed earlier [14, 16] allows one to obtain complete solutions of equations (5.9), (5.10). Letting $R = K - m(C)$ to be the resistance of microbes to the antimicrobial agent at concentration $C$, we can consider $R$ as the parameter distributed over the population. The distribution of $R$ can be easily computed if the distribution of concentration $C$ is known. The inhomogeneous model is of the form $dl(t,R)/dt = Rl(t,R)$ where $l(t,R)$ is the population density over the resistance $R$. Let $M_0(\lambda) = \int_A \exp(\lambda R) P(0,R) dR$ be the mgf of the initial distribution of resistance. Then the population size at moment $t$ $N(t) = N(0) M_0(t)$, and the current pdf of resistance $P(t,R) = P(0,R)\exp(tR)/M_0(t)$.

For example, if we suppose (as in [25]) that the initial distribution of the resistance is normal with a mean $m_0$, variance $\sigma_0^2$, and mgf $M_0(\lambda) = \exp(\lambda^2 \sigma_0^2 /2 + \lambda m_0)$, then the resistance distribution is also normal at any $t$ with the mean $E^t[R] = m_0 + \sigma_0^2 t$ and with the same variance $\sigma_0^2$, and $N(t) = N(0)\exp(t^2 \sigma_0^2 /2 + t m_0)$. Remark, that this supposition is not realistic as the resistance should be positive. If the initial distribution of the resistance is $\Gamma$ – distribution, $P(0,R) = s^k (R-\eta)^{k-1} \exp(-(R-\eta)s)/\Gamma(k)$, with mean $E^0[R] = \eta + k/s$, variance $\sigma_0^2 = k/s^2$, and the mgf $M_0(\lambda) = \exp(\lambda \eta)/(1-\lambda/s)^k$ for $\lambda < s$, then $R$ is also $\Gamma$ – distributed at any moment $t < s$ with mean $E^t[R] = \eta + k/(s-t)$ and variance $\sigma_t^2 = k/(s-t)^2$, and $N(t) = N(0)\exp(t\eta)/(1-t/s)^k$. The model "blows up" when $t = s$, i.e. the population size tends to infinity as $t \to s$ and hence the model is unrealistic.

The normal and $\Gamma$ – distributions are completely characterized by their mean and variance; it follows from the examples given above that the fate of the population can be dramatically different at the same initial mean and variance of the resistance.

More realistically, the actual initial distribution of the resistance should be concentrated in a finite interval; we can suppose that the initial distribution is uniform, Beta-distribution, or truncated exponential in that interval. In all these cases the model can be solved explicitly. For example, in the last case $P(0,R) = V\exp(-sR)$ where $0 \leq R \leq c = const$, $s$ is the distribution parameter, and $V = s/(1-\exp(-sc))$ is the normalization constant. Then the current population size $N(t)$ is defined by the formula $N(t) = N(0)(1-t/s)^{-1}(1-\exp(c(t-s)))/(1-\exp(-sc))$ and the current distribution of $R$ is the truncated exponential distribution with the parameter $s$-$t$ (see [16] for details).

Next, let us consider a more realistic logistic version (5.10) of the model. Corresponding inhomogeneous model reads

$$dl(t,C)/dt = l(t,C)(K(1-N(t)/B) - m(C)) \qquad (5.11)$$

where $l(t,C)$ is the microbial subpopulation under the pressure of antimicrobial agent with concentration $C$. The reproduction rate depends only on the total population size, so in order to solve this equation we only need to know the mgf of the initial distribution $P(0,C)$ (see ss.4.2, 4.3)). Let

$$M_0(\lambda,\delta) = \int_A \exp(\lambda + \delta m(C))P(0,C)dC = \exp(\lambda)E^0[\exp(\delta m)].$$ Then the escort system for auxiliary variables read (see (4.5)):

$ds/dt = -1, s(0) = 0$, hence $s(t) = -t$;

$$dq/dt = K(1 - N(0)\exp(q)E^0[\exp(-tm)]/B) = K(1 - A(t)\exp(q)), q(0) = 0. \qquad (5.12)$$

Here we denote $A(t) = N(0)E^0[\exp(-tm)]/B$. The function $A(t)$ is known at a given initial distribution and hence equation (5.12) can be solved (at least, numerically). The solution to model (5.11) is $l(t,C) = l(0,C)\exp(q(t) - tm(C))$; the total population size $N(t) = N(0)\exp(q(t))E^0[\exp(-tm)]$; the distribution of the agent concentration at moment $t$ is $P(t,C) = P(0,C)\exp(-tm(C))/E^0[\exp(-tm)]$.

The model and its solution are simplified if we consider the kill rate $m$ rather then the concentration $C$ as a distributed parameter. In this case the model reads

$$dl(t,m)/dt = l(t,m)(K(1-N(t)/B) - m). \qquad (5.13)$$

Let $M_0(\lambda) = E^0[\exp(\lambda m)]$ be the mgf of initial distribution of the kill rate $m$. Changing the variable, $z = \exp(q)$, we obtain the equation $dz/dt = zK(1 - zN(0)M_0(-t)/B)$, $z(0) = 1$. This equation can be solved at the given mgf $M_0$ of the initial distribution.

The solution to model (5.13) is now $l(t,m) = l(0,m)z(t)\exp(-tm)$; the total population size $N(t) = N(0)z(t)M_0(-t)$, and the distribution of the kill rate $P(t,m) = P(0,m)\exp(-tm)/M_0(-t)$.

**Example** 5.

Many particular models have the form of the inhomogeneous logistic equation

$$dl(t;\beta,\mu)/dt = l(t;\beta,\mu)[(\beta f_1(N(t)) - \mu f_2(N(t)))]. \tag{5.14}$$

(see, e.g., [2] and references therein). Let us give explicit formulas for the solution of this equation. Let $M_0(\lambda_1, \lambda_2) = E^0[\exp(\lambda_1\beta + \lambda_2\mu)]$ be the mgf of the joint initial distribution of $\beta$ and $\mu$. The escort system for auxiliary variables reads

$$dq_1/dt = f_1(N(0)M_0(q_1,q_2)), q_1(0) = 0; \tag{5.15}$$

$$dq_2/dt = -f_2(N(0)M_0(q_1,q_2)), q_2(0) = 0.$$

The solution to equation (5.14) is

$$l(t;\beta,\mu) = l(0;\beta,\mu)\exp(q_1(t)\beta + q_2(t)\mu), \tag{5.16}$$

the total population size is given by

$$N(t) = N(0)M_0(q_1(t), q_2(t))$$

and the current distribution is given by the formula

$$P(t;\beta,\mu) = P(0;\beta,\mu)\exp(q_1(t)\beta + q_2(t)\mu)/M_0(q_1(t),q_2(t)). \tag{5.17}$$

A particular case of equation (5.14)

$$dl(t;\beta,\mu)/dt = l(t;\beta,\mu)(\beta - \mu N(t)) \tag{5.18}$$

was studied in [1] for independent growth and mortality parameters, $\beta$ and $\mu$, uniformly distributed in the intervals $[a_1, b_1]$ and $[a_2, b_2]$ accordingly. The theorem of existence and uniqueness was established; it was also proven that the population concentrates asymptotically in the parametric point $[b_1, a_2]$ with the highest growth to mortality ratio. Remark that both conditions, the independence of the parameters and boundedness of their domains of values are essential for the last statement to be valid. For example, if $\beta = c\mu$ then the population does not concentrate in a parametric point but stays inhomogeneous indefinitely [16]. The second condition is discussed below.

Let us solve equation (5.18). The first equation of the escort system (5.15) reads $dq_1/dt = 1, q_1(0) = 0$, hence $q_1(t) = t$. So, the second equation is $dq_2/dt = -N(0)M_0(t, q_2)$, $q_2(0) = 0$.

Let the parameters $\beta \in [a_1, b_1]$ and $\mu \in [a_2, b_2]$ be independent and uniformly distributed at the initial moment, i.e. $P(0; \beta, \mu) = 1/((b_1 - a_1)(b_2 - a_2))$. The mgf of the uniform distribution in $[a,b]$ is $M_0(\lambda) = (\exp(\lambda b) - \exp(\lambda a))/(\lambda(b-a))$, hence

$$M_0(\lambda_1, \lambda_2) = E^0[\exp(\lambda_1 \beta + \lambda_2 \mu)] =$$
$$(\exp(\lambda_1 b_1) - \exp(\lambda_1 a_1))/(\lambda_1(b_1 - a_1))(\exp(\lambda_2 b_2) - \exp(\lambda_2 a_2))/(\lambda_2(b_2 - a_2)).$$

Then the auxiliary variable solves the equation

$$dq_2/dt = -N(0)M_0(t, q_2) =$$
$$-N(0)\frac{\exp(tb_1) - \exp(ta_1)}{t(b_1 - a_1)} \frac{\exp(q_2(t)b_2) - \exp(q_2(t)a_2)}{q_2(t)(b_2 - a_2)}.$$

The solution of equation (5.18) is then

$$l(t; \beta, \mu) = 1/((b_1 - a_1)(b_2 - a_2))\exp(t\beta + q_2(t)\mu).$$

Next, $P(t; \beta, \mu) = P^1(t; \beta)P^2(t; \mu)$ where

$$P^1(t; \beta) = \frac{\exp(t\beta)}{(b_1 - a_1)E^0[\exp(t\beta)]}, \quad P^2(t; \mu) = \frac{\exp(q_2(t)\mu)}{(b_2 - a_2)E^0[\exp(q_2(t)\mu)]}.$$

It is now easy to see (taking into account that $q_2(t) \to -\infty$ as $t \to \infty$) that $P_1(t; \beta)$ and $P_2(t; \mu)$ in course of time concentrate at points $b_1$ and $a_2$ correspondingly, in accordance with [1].

The following example shows that qualitatively different asymptotical behavior of the same equation is possible with another initial distribution. Let the positive parameters $\beta, \mu$ be independent again, and the initial distribution of both parameters be exponential, $P_i(x) = s_i \exp(-x s_i)$, with the mgf $M_i(\lambda) = 1/(1 - \lambda/s_i)$, $i = 1, 2$. Let us put $s_1 = T, s_2 = 1$, and $N(0) = 1$ for simplicity. Then $M_0(\lambda_1, \lambda_2) = \dfrac{1}{(1 - \lambda_1/T)(1 - \lambda_2)}$, and the auxiliary variable solves the equation

$$dq_2/dt = -N(0)M_0(t, q_2) = -\frac{1}{(1 - t/T)(1 - q_2)}, \quad q_2(0) = 0, \tag{5.19}$$

which can be easily integrated: $q_2(t) = 1 - \sqrt{1 - 2T\ln(1 - t/T)}$.

Now we can see that the global solution of equation (5.19) exists only at $t < T$ and $q_2(t) \to -\infty$ as $t \to T$. It means that the solution of the inhomogeneous logistic equation (5.18) with exponentially distributed parameters exists only up to the moment $t = T$. The model solution $l(t; \beta, \mu)$, the total population size and the parameter distributions for $t < T$ can be written down with the help of formulas (5.16)-(5.17). In particular, $N(t) = N(0)/\{(1 - t/T)\sqrt{1 - 2T\ln(1 - t/T)}\}$. As $t \to T$, the total population size tends to infinity and the population vanishes in any finite interval of values for both parameters, $\beta$ and $\mu$. Remark, that a similar phenomenon of "population explosion" at a certain time moment $T < \infty$ is realized for a wide class of $\Gamma-$ distributed parameters.

**Example 6**.

To conclude this section, let us demonstrate how to solve the inhomogeneous version of the well-known Ricker equation (see, e.g., [37], s.5.3)):

$$dl(t; \beta, \mu)/dt = l(t; \beta, \mu)[(\beta \exp(-cN(t))) - \mu]. \qquad (5.20)$$

Let $M_0(\lambda_1, \lambda_2)$ be the mgf of the joint initial distribution of $\beta$ and $\mu$. Then the escort system reads

$dq_2/dt = -1, q_2(0) = 0$, hence $q_2(t) = -t$;

$$dq_1/dt = \exp(-cN(0)M_0(q_1, -t)), q_1(0) = 0. \qquad (5.21)$$

Equation (5.21) can be solved at known $M_0$ and then the solution to (5.20) is equal to

$l(t; \beta, \mu) = l(0; \beta, \mu)\exp(q_1(t)\beta - t\mu)$; the total population size $N(t) = N(0)M_0(q_1(t), -t)$ and the system distribution $P(t; \beta, \mu) = P(0; \beta, \mu)\exp(q_1(t)\beta - t\mu)/M_0(q_1(t), -t)$.

For example, let the parameters $\beta$ and $\mu$ be independent and exponentially distributed in $[0, \infty)$ with the means $s_1$ and $s_2$ at the initial instant. Then $M_0(q, -t) = s_1 s_2/((s_1 - q)(s_2 + t))$, and equation (5.21) for the auxiliary variable reads

$dq_1/dt = \exp(-cN(0)s_1 s_2/((s_1 - q_1)(s_2 + t)), q_1(0) = 0$.

This equation has a stable state $q_1 = s_1$. As $t \to \infty$, $q_1(t) \to s_1$, the total population size tends to infinity and the population density concentrates at the value $\mu = 0$ of the parameter $\mu$ and vanishes in any finite interval of values of the parameter $\beta$.

**Discussion**

Mathematical theory of selection has a long history; R. Fisher, S. Wright, J. Haldane were its father-founders. G. Price was the first who tried to find a general formulation of selection that could be applied to any (not only biological) problem and to develop a formal general theory. He hoped that the concept of selection proposed in his paper [32], which was published only in 1995, "will contribute to the future development of 'selection theory' as helpfully as Hartley's concept of information contributed to Shannon's communication theory. … Many scientists must have felt surprise to find that at so late date there had still remained an opportunity to develop so fundamental a scientific area. Perhaps a similar opportunity exists today in respect to selection theory".

The Price equation was an outstanding contribution to the future theory; its particular cases are the Fisher fundamental theorem and the covariance equation. The Price equation is universally applicable to any selection systems at any instant independently from the underlying mathematical model and its specific dynamics (see, e.g., [33], ch.6 for details) because this equation is a mathematical *identity*. It is a reason for the theoretical universality and practical unavailing ("dynamical insufficiency") of the Price equation.

The Haldane optimal principle can be considered as one of the first general assertions about selection systems; it describes the asymptotical behavior of a population composition. This principle was generalized in [35], ch.3 for models with discrete time and in [7] (Appendix) for models with continuous time. The authors developed an abstract theory of systems with inheritance and applied it to some problems of mathematical biology (see, e.g., monograph [9] and the survey [8]). The "resampling down" of the initial variety demonstrates the qualitative effects of "natural selection" as $t \to \infty$. The theory gives a complete description of the support of the limit distributions but the dynamics of systems on a finite time interval is out of the scope of this theory.

Thus, at the present time the mathematical theory of selection has general frameworks for mathematical modeling (the systems with inheritance, replication equations, selection systems), some fundamental assertions (like the Haldane principle, the Fisher fundamental theorem of selection and the Price equations) and many particular models of inhomogeneous populations and communities. Most of these models have a form of many- or infinite dimensional differential or difference systems of equations. Some theorems of existence and uniqueness and asymptotic behavior of solutions to particular classes of such equations were established (see, e.g., [1, 8, 28]), but no general methods for solving the models analytically are known, except for numerical investigation.

In this paper we suggest a general approach to a wide class of self-regulated selection systems. The main evolutionary forces in an evolving system are selection, mutation and random drift; it should be noticed that our approach explicitly examines only selection. The developed theory allows one to reduce the complex inhomogeneous models to the "escort systems" of ODEs that, in many cases, can be investigated analytically. Notice that even if the analytical solution of the escort system is not available, numerical solving the Cauchy problem for a system of ODE is much simpler than studying the initial problem numerically. It allows us to compute (in many cases, explicitly) the evolution of distributions and all statistical characteristics of interest of the initial selection system. Similar approach to discrete-time models was developed in [19]. The considered examples show how different the global dynamics of a selection system can be depending on the initial distribution even for the same dynamical model.

We have systematically applied our approach to inhomogeneous Malthusian and logistic equations; explicit solutions to different models of these types used in the literature were derived. Analytical solutions of the considered models can provide new biological insights beyond the computer simulations performed in the original papers. We believe that derived explicit solutions may be helpful and necessary in order to be able to completely study corresponding models, which belong to different areas of mathematical biology; however, that is out of the scope of this paper. Applications to forest ecology modeling [15], global demography [16], cancer modeling [18], and epidemics in heterogeneous populations [26] were published recently. It is our hope that the developed approach has potential for different applications these and others areas of science.

**Acknowledgement.** This research was supported [in part] by the Intramural Research Program of the NIH, National Library of Medicine. The author thanks Dr. E. Koonin, Dr. F. Berezovsky, Dr. A. Novozhilov and anonymous reviewer for valuable comments.

**Appendix**

Proof of Theorem 1

Let $\{\mathbf{q}(t), \mathbf{s}(t)\}$ be a solution of Cauchy problem (3.5)- (3.6); denote for instant

$$K*_t(\mathbf{a}) = \exp(\sum_{i=1}^{n} q_i(t)\varphi_i(\mathbf{a}) + \sum_{k=1}^{m} s_k(t)\psi_k(\mathbf{a})),$$

$$l*(t,\mathbf{a})/dt = l(0,\mathbf{a})K*_t(\mathbf{a}),$$

$$N*(t) = N(0)\Phi(1;\mathbf{q}(t),\mathbf{s}(t)),$$

$$G*_i(t) = N(0)\Phi(g_i;\mathbf{q}(t),\mathbf{s}(t)),$$

$$H*_i(t) = \Phi(h_i;\mathbf{q}(t),\mathbf{s}(t))/\Phi(1;\mathbf{q}(t),\mathbf{s}(t)),$$

$$F*(t,\mathbf{a}) = \sum_{i=1}^{n} u_i(t,G*_i)\varphi_i(\mathbf{a}) + \sum_{k=1}^{m} v_k(t,H*_k)\psi_k(\mathbf{a}),$$

$$P*(t,\mathbf{a}) = P*(0,\mathbf{a})K*_t(\mathbf{a})/\Phi(1;\mathbf{q}(t),\mathbf{s}(t)).$$

Then $dl*(t,\mathbf{a})/dt = l(0,\mathbf{a})K*_t(\mathbf{a})F*_t(\mathbf{a}) = l*(t,\mathbf{a})F*_t(\mathbf{a})$. Next,

$$\int_A l*(t,\mathbf{a})d\mathbf{a} = \int_A l(0,\mathbf{a})K*_t(\mathbf{a})d\mathbf{a} = N(0)\Phi(1;\mathbf{q}(t),\mathbf{s}(t)) = N*(t);$$

$$\int_A g(\mathbf{a})l*(t,\mathbf{a})d\mathbf{a} = \int_A g(\mathbf{a})K*_t(\mathbf{a})l(0,\mathbf{a})d\mathbf{a} = N(0)\Phi(g;\mathbf{q}(t),\mathbf{s}(t)) = G*(t);$$

$$\int_A h(\mathbf{a})P*(t,\mathbf{a})d\mathbf{a} = \int_A h(\mathbf{a})K*_t(\mathbf{a})P(0,\mathbf{a})/\Phi(1;\mathbf{q}(t),\mathbf{s}(t))d\mathbf{a} =$$

$$\Phi(h;\mathbf{q}(t),\mathbf{s}(t))/\Phi(1;\mathbf{q}(t),\mathbf{s}(t)) = H*(t).$$

It means that functions $l*(t,\mathbf{a}), N*(t), G*_i(t), H*_k(t)$ satisfy system (3.5)-(3.6).

Conversely, let $l(t,\mathbf{a}), N(t), G_i(t), H_k(t)$ solve system (3.3) for $t \in [0,T)$. For now, define the functions $q_i*(t), s_k*(t)$, $i = 1,...n, k = 1,...m$ by relations:

$$q_i*(t) = \int_0^t u_i(x,G_i(x))dx, \text{ and } s_k*(t) = \int_0^t v_k(x,H_k(x))dx; \text{ then}$$

$$dl(t,\mathbf{a})/l(t,\mathbf{a}) = \sum_{i=1}^{n} \varphi_i(\mathbf{a})dq*_i(t) + \sum_{k=1}^{m} \psi_k(\mathbf{a})ds*_k(t), \text{ and}$$

$$l(t,\mathbf{a}) = l(0,\mathbf{a})\exp(\sum_{i=1}^{n} \varphi_i(\mathbf{a})dq*_i(t) + \sum_{k=1}^{m} \psi_k(\mathbf{a})ds*_k(t)) \text{ for all } t \in [0,T).$$

Hence, $N(t) = \int_A l(t,\mathbf{a})d\mathbf{a} = \int_A l(0,\mathbf{a})\exp(\sum_{i=1}^{n} \varphi_i(\mathbf{a})q*_i(t) + \sum_{k=1}^{m} \psi_k(\mathbf{a})s*_k(t))d\mathbf{a} =$

$N(0)\Phi(1;\mathbf{q}*(t),\mathbf{s}*(t));$

$$G(t) = \int_A g(\mathbf{a})l(t,\mathbf{a})d\mathbf{a} = \int_A g(\mathbf{a})\exp(\sum_{i=1}^{n}\varphi_i(\mathbf{a})q*_i(t) + \sum_{k=1}^{m}\psi_k(\mathbf{a})s*_k(t))l(0,\mathbf{a})d\mathbf{a} =$$

$$N(0)\Phi(g;\mathbf{q}*(t),\mathbf{s}*(t));$$

$$H(t) = \int_A h(\mathbf{a})l(t,\mathbf{a})d\mathbf{a}/N(t) = N(0)M(h;\mathbf{q}*(t),\mathbf{s}*(t))/N(t) =$$
$$\Phi(h;\mathbf{q}*(t),\mathbf{s}*(t))/\Phi(1;\mathbf{q}*(t),\mathbf{s}*(t)).$$

From the definition, $\{\mathbf{q}*(t),\mathbf{s}*(t)\}$ is the solution of Cauchy problem (3.5)- (3.6) for $t \in [0,T)$. Theorem is proven.